\documentclass[aip,jcp,graphicx]{revtex4-1}

\usepackage{graphicx}
\usepackage[T1]{fontenc}
\usepackage{xcolor}
\usepackage{amssymb}
\usepackage{amsmath}
\usepackage{blkarray}
\usepackage{array,booktabs,tabularx}
\usepackage{multirow}
\usepackage{natbib}
\usepackage{url}

\newcolumntype{C}{>{\centering\arraybackslash}X}

\begin{document}


\title{Chemical Potential Calculations in Non-Homogeneous Liquids}



\author{C. Perego}
\email[]{perego@mpip-mainz.mpg.de}
\affiliation{  Department of Polymer Theory, Max-Planck Institute for Polymer Research, Ackermannweg 10, D-55128 Mainz (Germany)}
\affiliation{  Department of Chemistry and Applied Biosciences, ETH Zurich, c/o USI Campus, Via Giuseppe Buffi 13, CH-6900 Lugano (Switzerland)} 
\affiliation{  Universit\`a della Svizzera italiana, Institute of Computational Science, National Center for Computational Design and Discovery of Novel Materials MARVEL, Universit\`a della Svizzera italiana, Via Giuseppe Buffi 13, CH-6900 Lugano (Switzerland)}

\author{O. Valsson }%
\affiliation{  Department of Polymer Theory, Max-Planck Institute for Polymer Research, Ackermannweg 10, D-55128 Mainz (Germany)}
\affiliation{  Department of Chemistry and Applied Biosciences, ETH Zurich, c/o USI Campus, Via Giuseppe Buffi 13, CH-6900 Lugano (Switzerland)} 
\affiliation{  Universit\`a della Svizzera italiana, Institute of Computational Science, National Center for Computational Design and Discovery of Novel Materials MARVEL, Universit\`a della Svizzera italiana, Via Giuseppe Buffi 13, CH-6900 Lugano (Switzerland)}

\author{M. Parrinello}
%
\affiliation{  Department of Chemistry and Applied Biosciences, ETH Zurich, c/o USI Campus, Via Giuseppe Buffi 13, CH-6900 Lugano (Switzerland)} 
\affiliation{  Universit\`a della Svizzera italiana, Institute of Computational Science, National Center for Computational Design and Discovery of Novel Materials MARVEL, Universit\`a della Svizzera italiana, Via Giuseppe Buffi 13, CH-6900 Lugano (Switzerland)}




\begin{abstract}
The numerical computation of chemical potential in dense, non-homogeneous fluids is a key problem in the study of confined fluids thermodynamics. To this day several methods have been proposed, however there is still need for a robust technique, capable of obtaining accurate estimates at large average densities. A widely established technique is the Widom insertion method, that computes the chemical potential by sampling the energy of insertion of a test particle. Non-homogeneity is accounted for by assigning a density dependent weight to the insertion points. However, in dense systems, the poor sampling of the insertion energy is a source of inefficiency, hampering a reliable convergence. 

We have recently presented a new technique for the chemical potential calculation in homogeneous fluids. This novel method enhances the sampling of the insertion energy via Well-Tempered Metadynamics, reaching accurate estimates at very large densities. In this paper we extend the technique to the case of non-homogeneous fluids.
The method is successfully tested on a confined Lennard-Jones fluid. In particular we show that, thanks to the improved sampling, our technique does not suffer from a systematic error that affects the classic Widom method for non-homogeneous fluids, providing a precise and accurate result.
\end{abstract}

\maketitle

\section{Introduction}\label{intro}

Chemical potential regulates a wide range of chemico-physical processes like phase equilibrium or chemical reactions\cite{JobEJP2006,BaierleinAJP2001}. However, its calculation in atomistic simulations is not without difficulties. More than fifty years ago Widom proposed a practical way of computing chemical potential in Monte Carlo or Molecular Dynamics (MD) simulations\cite{WidomJCP1963}. This method, also known as test particle insertion method, consists in periodically sampling the insertion energy of an extra particle during a running simulation. For fluids at low or moderate densities Widom's method is simple and efficient, but in dense fluids the probability of inserting a test particle is vanishingly small and the method becomes impractical. Later, in an attempt at overcoming this problem, Bennet proposed to combine insertion and deletion computations\cite{BennettJCP1976} in an optimal way. This led to improved accuracy\cite{LuJCP2001a,LuJCP2001b} but the limitations in the case of dense fluids still persisted. After these pioneering approaches several, more complex attempts at extending chemical potential calculations into the high density region have been presented, e.g.~exploiting Free Energy Perturbation (FEP)\cite{ShingMP1982,LuJCP1999}, Thermodynamic Integration (TI)\cite{BeutlerJCP1995}, or alternative approaches (see e.g.~Refs.~\onlinecite{DingJCP1993,DelgadoJCP2005,BoulougourisJCED2010,MooreJCP2011,AgarwalJCP2014}). Review papers comparing some of these methods are available\cite{KofkeMP1997,LuJCP2003,ShirtsJCP2005,DalyCPC2012}, but the accuracy and efficiency of these techniques strongly depend on the physical attributes of the system under study. For example, most of the cited methods deal with uniform systems, whereas many applications, such as the study of phase coexistence \cite{HeffelfingerMP1987,MiyaharaJCP2013} or nanoconfined fluids \cite{JiangNL2004} require the chemical potential calculation in a non-homogeneous system. Widom test particle approach can be extended to non-homogeneous fluids \cite{WidomJSP1978,PowlesJCP1994} but also alternative methods have been proposed, exploiting Gibbs ensemble \cite{PanagiotopoulosMP1987}, Grand-Canonical Monte Carlo\cite{PetersonMP1987,PapadopolouJCP1992} or introducing an auxiliary cell coupled with the non-homogeneous fluid \cite{MiyaharaJCP1997,NeimarkJCP2005}.

Very recently we have proposed an alternative MD approach for computing the chemical potential of a dense fluid that has proven to be rather efficient\cite{PeregoEPJ2016}. The method is based on the identification of an appropriate Collective Variable (CV) that couples to those fluid configurations that more easily allow the insertion of a test particle. The fluctuations of this CV are then enhanced so as to favor successful insertions. In our previous work we assumed that the fluid was uniform. Here, stimulated by the interest in nano-fluidics and confined systems  we lift this limitation and extend the ideas of Ref.~\onlinecite{PeregoEPJ2016} to non-uniform systems.

\section{Theory and Methods}\label{methods}

For simplicity we restrict ourselves to the study of a single component fluid system, at constant volume $V$ and temperature $T$. Extension to multi-component systems is straightforward. The chemical potential $\mu$ can be expressed as the derivative of the Helmoltz free energy $F$ with respect to the number of particles $N$. We shall approximate this derivative with the finite difference:
\begin{equation}
\label{mudef}
\mu=\frac{\partial F(V,T,N)}{\partial N}\simeq F(V,T,N+1)-F(V,T,N).
\end{equation}
Finite size corrections to this formula have been considered\cite{SmitJPCM1989,HongJCP2012}, but they are not relevant in the present context and will be neglected.
We separate from $\mu$ the ideal gas term $\mu^{\mathrm{id}}$:
\begin{equation}
\label{mudef}
\mu=\mu^{\mathrm{id}}+\mu^{\mathrm{ex}},
\end{equation}
where:
\begin{equation}
\label{muid}
\mu^{\mathrm{id}}=-\beta^{-1}\log\left[\frac{V}{\Lambda^{3}(N+1)}\right],
\end{equation}
in which $\beta=(k_{\mathrm{B}}T)^{-1}$ is the inverse temperature and $\Lambda$ is de Broglie wavelength. The excess term $\mu^{\mathrm{ex}}$ is given by:
\begin{equation}
\label{muex}
\mu^{\mathrm{ex}}=-\beta^{-1}\log\left[\frac{1}{V}\frac{Z_{N+1}}{Z_{N}}\right],
\end{equation}
where $Z_{N}$ is the configurational partition function of the $N$ particle system:
\begin{equation}
\label{Z}
Z_{N}=\int\exp\left[-\beta U(\mathbf{R})\right]\,\mathrm{d}\mathbf{R},
\end{equation}
$U$ is the potential energy and $\mathbf{R}$ the atomic coordinates of the system.
One can then write:
\begin{equation}
\label{ZoZ}
\frac{Z_{N+1}}{Z_{N}}=\int\left\langle\exp\left[-\beta \Delta U(\mathbf{R}^{*};\mathbf{R})\right]\right\rangle_{N}\mathrm{d}\mathbf{R}^{*},
\end{equation}
in which $\mathbf{R}^{*}$ is the coordinate vector of the extra particle in the $N+1$ partition function, $\langle\ldots\rangle_{N}$ is the ensemble average of the $N$-atoms system and
\begin{equation}
\label{DU}
\Delta U(\mathbf{R}^{*};\mathbf{R})=U(\mathbf{R}^{*};\mathbf{R})-U(\mathbf{R})
\end{equation}
is the energy of insertion of a particle at $\mathbf{R}^{*}$.

Let us now briefly recall the method we have proposed for homogeneous systems, presented in Ref.~\onlinecite{PeregoEPJ2016}. First we make use of the fact that, in homogeneous fluids, $\langle\exp\left[-\beta \Delta U(\mathbf{R}^{*};\mathbf{R})\right]\rangle_{N}$ does not depend on $\mathbf{R}^{*}$, so that: 
\begin{equation}
\label{muh}
\mu^{\mathrm{ex}}=-\beta^{-1}\log\left\langle\exp\left[-\beta \Delta U(\mathbf{R}^{*};\mathbf{R})\right]\right\rangle_{N},
\end{equation}
and then introduce the CV:
\begin{equation}
\label{sform}
s_{\mathrm{h}}(\mathbf{R})=-\beta^{-1}\log\left[\frac{1}{M}\sum_{i=1}^{M}\exp\left(-\beta\Delta U(\mathbf{R}^{*}_{i};\mathbf{R})\right)\right],
\end{equation}
where we have added the subscript $\mathrm{h}$ to distinguish this CV from the general version valid also for non-homogeneous systems, to be introduced later. 
The CV $s_{\mathrm{h}}$ can be viewed as a generalized insertion energy, collecting the contribution of $M$ fixed insertion points $\mathbf{R}^{*}_{i}$ that are defined before the sampling. Note that if a single $\mathbf{R}^{*}$ is considered then $s_{\mathrm{h}}=\Delta U(\mathbf{R}^{*};\mathbf{R})$. 

Combining Eqs.~\ref{muex}, \ref{ZoZ} and \ref{sform} we can rewrite $\mu^{\mathrm{ex}}$ in terms of $s_{\mathrm{h}}$ as:
\begin{equation}
\label{muofs}
\mu^{\mathrm{ex}}=-\beta^{-1}\log\langle\mathrm{e}^{-\beta s_{\mathrm{h}}}\rangle_{N}=-\beta^{-1}\log\int\mathrm{e}^{-\beta s_{\mathrm{h}}}p_{\mathrm{h}}(s_{\mathrm{h}})\,\mathrm{d}s_{\mathrm{h}},
\end{equation}
where $p_{\mathrm{h}}(s_{\mathrm{h}})$ is the canonical probability distribution of $s_{\mathrm{h}}$. The accurate estimate of Eq.~\ref{muofs} requires a thorough sampling of the negative $s_{\mathrm{h}}$ region, where the integrand $\mathrm{e}^{-\beta s_{\mathrm{h}}}p_{\mathrm{h}}(s_{\mathrm{h}})$ peaks. To this purpose we employ the well-known Well-Tempered (WT) metadynamics\cite{BarducciPRL2008}, in which a history-dependent bias potential is constructed, to enhance the statistical fluctuations along some chosen degrees of freedom of the system, namely the CVs. This generates a biased ensemble, in which the tails of the distribution are more accurately sampled. The correct canonical distribution can be then reconstructed starting from this biased ensemble (for more details see e.g.~Ref.~\onlinecite{ValssonRev2015}). In our case $s_{\mathrm{h}}$ can be used as biasing CV in a metadynamics procedure, to drive the sampling in the region of interest for Eq.~\ref{muofs}. This choice results in an accurate computation of $\mu^{\mathrm{ex}}$ that, for dense systems, surpasses Widom's method in efficiency\cite{PeregoEPJ2016}.

We now extend this approach to the case of non-homogeneous systems, where the translational invariance does not hold. Following Ref.~\onlinecite{WidomJSP1978}, we first note that the average density of the $N+1$ particle system at $\mathbf{R}^{*}$ is given by:
\begin{equation}
\label{Z}
\rho_{N+1}(\mathbf{R}^{*})=\frac{N+1}{Z_{N+1}}\int\exp\left[-\beta U(\mathbf{R}^{*};\mathbf{R})\right]\,\mathrm{d}\mathbf{R},
\end{equation}
which, with simple manipulations, can be written as:
\begin{equation}
\label{Z}
\rho_{N+1}(\mathbf{R}^{*})=(N+1)\frac{Z_{N}}{Z_{N+1}}\left\langle\exp\left[-\beta \Delta U(\mathbf{R}^{*};\mathbf{R})\right]\right\rangle_{N},
\end{equation}
leading to the relation:
\begin{equation}
\label{widomnon}
\frac{Z_{N+1}}{Z_{N}}=\frac{N+1}{\rho_{N+1}(\mathbf{R}^{*})}\left\langle\exp\left[-\beta \Delta U(\mathbf{R}^{*};\mathbf{R})\right]\right\rangle_{N},
\end{equation}
that shows how the chemical potential does not depend on the local density in $\mathbf{R}^{*}$. In Eq.~(\ref{widomnon}) the choice of $\mathbf{R}^{*}$ is arbitrary, and therefore we can introduce the CV:
\begin{eqnarray}
\label{snon}
s(\mathbf{R})&=&-\beta^{-1}\log\left[\frac{1}{M}\sum_{i=1}^{M}\frac{N+1}{V\rho_{N+1}(\mathbf{R}_{i}^{*})}\exp\left(-\beta\Delta U(\mathbf{R}^{*}_{i};\mathbf{R})\right)\right]\\
\label{snon2}&=&-\beta^{-1}\log\left[\frac{1}{M}\sum_{i=1}^{M}\exp\left(-\beta\Delta \tilde{U}(\mathbf{R}^{*}_{i};\mathbf{R})\right)\right].
\end{eqnarray}
Which is a generalization of Eq.~\ref{sform} to non-homogeneous systems. In the last equality (Eq.~\ref{snon2}) we have introduced an effective insertion energy that contains the local density information:
\begin{equation}
\label{effu}
\Delta\tilde{U}(\mathbf{R}^{*}_{i};\mathbf{R})=\Delta U(\mathbf{R}^{*}_{i};\mathbf{R})+\beta^{-1}\log\left(\frac{\rho_{N+1}(\mathbf{R}_{i}^{*})}{\rho_{0,N+1}}\right),
\end{equation}
in which $\rho_{0,N+1}=(N+1)/V$ is the average density of the $N+1$ system. 
As in Ref.~\onlinecite{PeregoEPJ2016}, the $M$ insertion points $\mathbf{R}_{i}^{*}$ are chosen before the metadynamics run. 
In principle the use of Eq.~(\ref{snon}) would require knowledge of $\rho_{N+1}(\mathbf{R}_{i}^{*})$. Although this quantity could be easily computed by performing a separate simulation in an $N+1$ particle system, for large $N$ the difference between $\rho_{N+1}$ and  $\rho_{N}$ is so small that we shall neglect it in the following and use $\rho_{N}(\mathbf{R}_{i}^{*})$. For simplicity we will remove the subscript and $\rho\equiv\rho_{N}$ from now on. 
This way of dealing with non-homogeneous systems mirrors what is done in the non-homogeneous extension of Widom's method \cite{WidomJSP1978,PowlesJCP1994}, where Eq.~(\ref{widomnon}) is used to assign a weight proportional to $\rho(\mathbf{R}_{i}^{*})^{-1}$ to each insertion point.

Using Eq.~(\ref{widomnon}) we can calculate $\mu^{\mathrm{ex}}$ in term of $s$ as:
\begin{equation}
\label{muofs2}
\mu^{\mathrm{ex}}=-\beta^{-1}\log\int\mathrm{e}^{-\beta s}p(s)\,\mathrm{d}s,
\end{equation}
which has a look identical to that of Eq.~(\ref{muofs}). 
Thus, as for the homogeneous case, WT metadynamics can be used to enhance the fluctuations of $s$ and collect the sampling required for the accurate calculation of $\mu^{\mathrm{ex}}$. 

In the following we report the results obtained using WT metadynamics, as it is nowadays a well-established technique. However we underline that alternative biasing techniques can be also employed. To prove this we have also applied our method enhancing the fluctuations via the recent Variationally Enhanced Sampling (VES) technique\cite{ValssonPRL2014}. The results of these calculations are reported in the Supplementary Material (SM).

\section{Test Case and Calculation Setup}\label{Setup}

\begin{figure}[] 
\centering \includegraphics{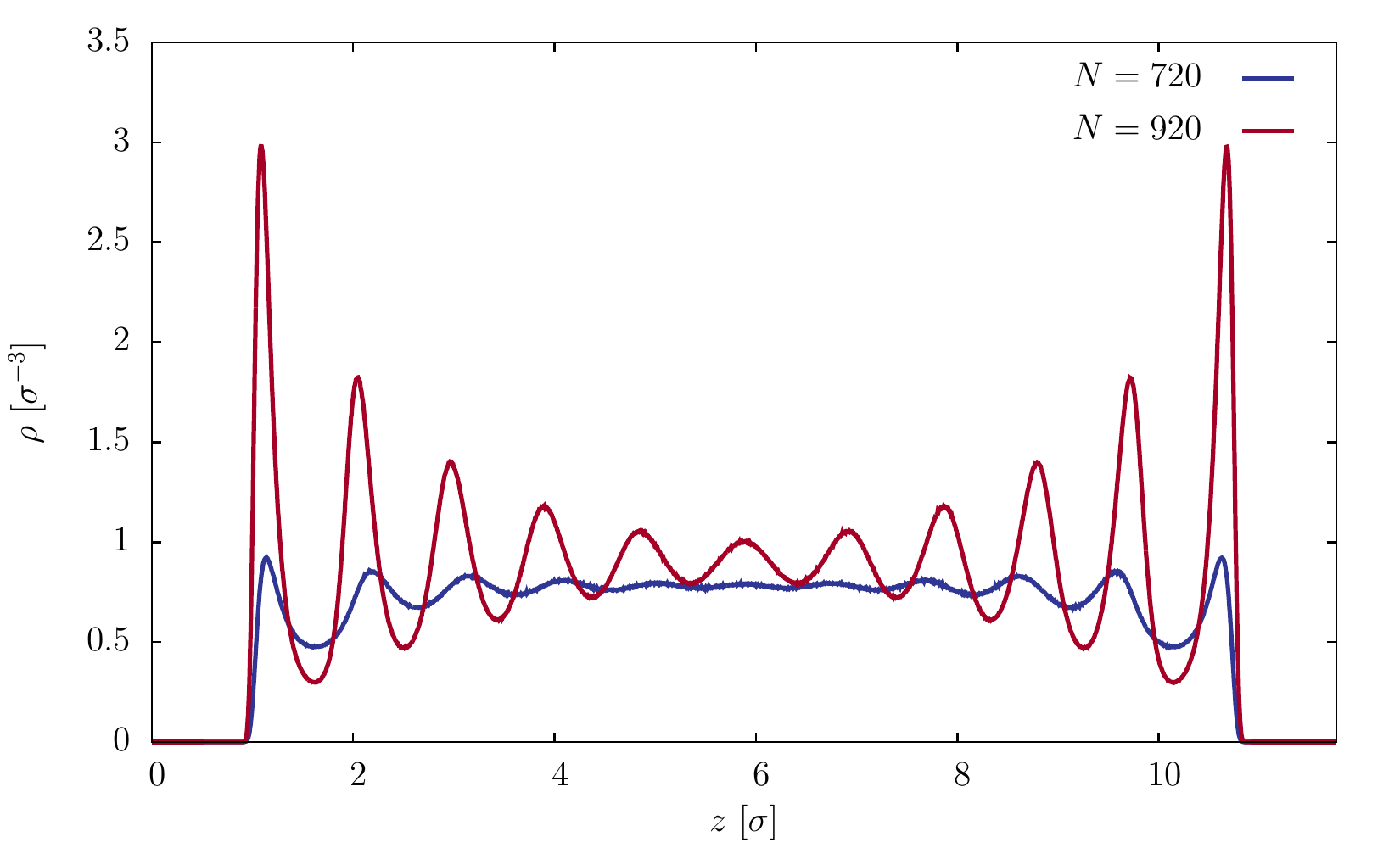} 
\caption{\label{dens} The number density profile $\rho$ as a function of the $z$ coordinate for the two simulated systems, with $N=720$ (blue curve) and $N=920$ (red curve) respectively.}\end{figure}

We study a single component Lennard-Jones (LJ) fluid, for which all the computed quantities are indicated in LJ units (energies are in units of $\epsilon$ and lengths in units of $\sigma$). The atoms interact via a LJ potential truncated and shifted at $r_{c}=2.5$. Two systems are considered, one with $N=720$ atoms, and the other with $N=920$ atoms. In both cases the temperature is $T=0.7$ and the simulation cell has the same size: $L_{x}=L_{y}=10.0$ along the $x$ and $y$ axes, and $L_{z}=11.762$ along the $z$ axis.  Both systems are in the liquid phase but at different average densities.
Periodic boundary conditions are imposed along $x$ and $y$, while the atoms are confined along the $z$ dimension by two LJ potential walls located at the boundaries of the box, mimicking fluid confinement in a slit pore. Each wall is represented by a LJ potential $U(z)$, with $\epsilon_{w}=1$ and $\sigma_{w}=1$, shifted to 0 at $z=L_{z}$. The total external potential has the following form:
\begin{equation}
\label{Uwalls}
U_{\mathrm{walls}}=U(z)+U(L_{z}-z).
\end{equation}
As shown in Fig.~\ref{dens}, the external walls induce a non-homogeneous average density profile, which reflects the combination of the attractive external potential with the liquid structure.
The system phase space is sampled using the LAMMPS\cite{PlimptonJCP1995,Lammpsurl} MD engine, linked to a private version of PLUMED 2\cite{TribelloCPC2014} for Metadynamics and VES calculations. MD is performed in the NVT ensemble, using the stochastic velocity rescaling thermostat\cite{BussiJCP2007}. The integration timestep is chosen to be $\Delta t = 10^{-3}$. 

The WT metadynamics calculations are performed depositing Gaussians of initial height $1.2$ and width $1.0$ every $\Delta\tau=500\,\Delta t$, with biasing factor $\gamma=30$ (see Ref.~\onlinecite{PeregoEPJ2016} for an overview on WT metadynamics theory in the same context). Before the production runs, we have tested different possibilities for the choice of the $M$ insertion points in Eq.~\ref{snon}, comparing the outcome of using regular grids, random uniform distributions, and distributions both directly and inversely proportional to $\rho$. For the systems studied we have found no significant differences among the grids, therefore we have decided to randomly choose $M$ insertion points in the system volume, avoiding the region close to the external walls, where the insertion probability is negligible (see Fig.~\ref{dens}). However, we cannot exclude that for other systems a different choice could be more appropriate. 

As already shown in the homogeneous fluid case~\cite{PeregoEPJ2016}, a regularized CV $s^{\mathrm{r}}$ needs to be defined, in order to prevent singularities in the computation of the bias forces:
\begin{equation}
\label{srnon}
s^{\mathrm{r}}(\mathbf{R})=-\beta^{-1}\log\left[\frac{1}{M}\sum_{i=1}^{M}\frac{\rho_{0,N+1}}{\rho(\mathbf{R}_{i}^{*})}\exp\left(-\beta\Delta U^{\mathrm{r}}(\mathbf{R}^{*}_{i};\mathbf{R})\right)\right],
\end{equation}
in which $U^{\mathrm{r}}$ is a regularized interaction potential. As done in Ref.~\onlinecite{PeregoEPJ2016}, we define $U^{\mathrm{r}}$ to be a LJ potential with quadratic soft-core for $r<r_{0}=0.881$. The metadynamics bias is then applied to $s^{\mathrm{r}}$, while the canonical $s$ distribution can be reconstructed after the simulation via the reweighting procedure, as explained in Ref.~\onlinecite{ValssonRev2015}.

Two kinds of metadynamics calculations are performed, type I and type II.
In type I the density profile $\rho(z)$ resulting from an unbiased trajectory is used to define both $s$ and $s^{\mathrm{r}}$ according to Eqs.~\ref{snon} and \ref{srnon} respectively.
This requires the generation of an extra unbiased trajectory to evaluate $\rho$. We underline that, while in general the calculation of $\rho$ is not demanding, its accuracy is relevant for the chemical potential calculation in non-homogeneous fluids (see Eq.~\ref{widomnon}). This holds in particular in our metadynamics method, in which the insertion points are fixed, and are just a few. Therefore the computation of $\rho$ might require a non-negligible extra-sampling. 
To avoid this further cost, in type II metadynamics the average density $N/V$ is used instead of $\rho$ to define the driving variable $s^{\mathrm{r}}$. The biased sampling is thus generated without prior knowledge of $\rho$, which is then computed from the biased trajectory during post-processing. Once $\rho$ is known $s$ is defined, and its distribution is evaluated via reweighting. In all calculations $\rho$ is computed with an histogram bin width of $\Delta z=5.0\times10^{-3}$.

We compare the results of our metadynamics method with both Widom and TI calculations.
For the Widom method we perform unbiased MD runs and compute the insertion energies of $M_{\mathrm{w}}$ randomly generated positions every $\Delta\tau$. The insertion points are uniformly distributed over the simulation box (as in metadynamics calculations, the region where the external potential diverges is avoided).

For the TI calculations we simulate the MD of an $N+1$ system, in which the $N+1$-th atom interaction with the system (the $N$ atoms and the confining potential) is gradually turned on through an integration parameter $\lambda$ that ranges from 0 (no interaction) to 1 (full interaction). 
To avoid singularities and instabilities the $N+1$-th atom interacts with the others via the soft-core potential introduced in Ref.~\onlinecite{BeutlerCPL1994}. The external potential action on the $N+1$-th is turned on by setting $\epsilon_{w}=\lambda^{2}$.
According to TI theory the free-energy difference between $\lambda=0$ and $\lambda=1$ is given by (see e.g.~Ref.~\onlinecite{MezeiANYAS1986}):
\begin{equation}
\label{tint}
\Delta F = \int_{0}^{1}\,\mathrm{d}\lambda'\left\langle\frac{\partial U}{\partial \lambda}\right\rangle_{\lambda=\lambda'} = \mu^{\mathrm{ex}},
\end{equation}
in which $U$ and $\langle\ldots\rangle_{\lambda=\lambda'}$ are respectively the potential energy and the ensemble average of the $N+1$ system with $\lambda=\lambda'$. As Eq.~\ref{tint} indicates, this free energy difference corresponds to the excess chemical potential of the system.
To approximate the integral in Eq.~\ref{tint} in our TI runs $n=40$ values $\lambda_{i}=i/n$ of the integration parameter are simulated, where $i=1\ldots n$. For every $\lambda_{i}$ the derivative of the potential energy with respect to $\lambda$ is sampled every $\Delta\tau$ to compute the ensemble average. When all the $\langle\partial U/\partial \lambda\rangle_{\lambda=\lambda_{i}}$ have been computed the integral is evaluated using the trapezoidal rule.

\section{Results}\label{Results}

\begin{figure}[] 
\centering \includegraphics{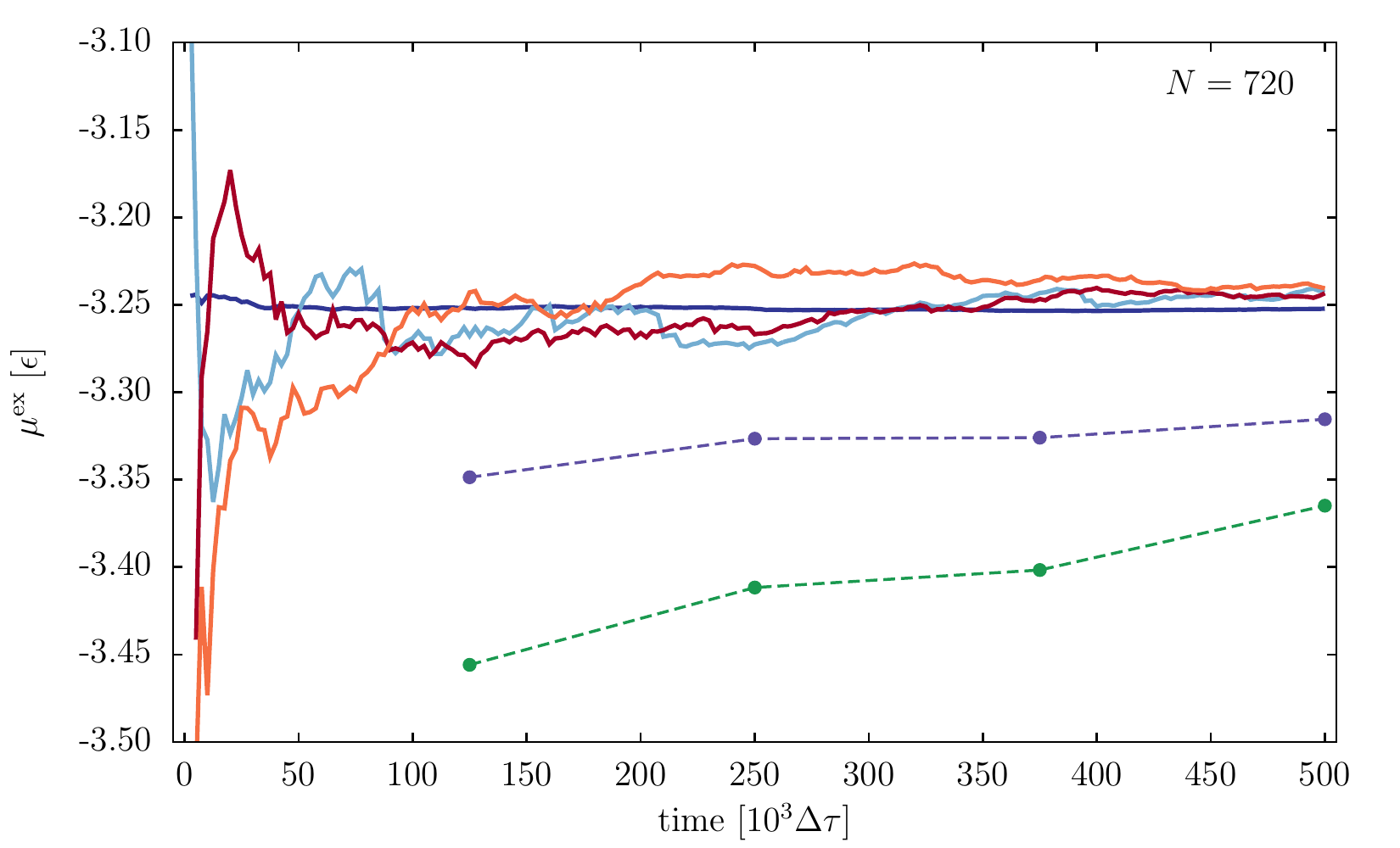} 
\caption{\label{mu1} $\mu^{\mathrm{ex}}$ as a function of the MD time for the $N=720$ system. The solid lines represent Widom results obtained with $M_{\mathrm{w}}=27000$ (blue) and  $M_{\mathrm{w}}=64$ (light blue), and metadynamics results with $M=64$ of type I (red) and II (orange). The dots represent TI-a (green) and TI-b (purple) results. The final results, and respective statistical errors, are indicated in Tab.~\ref{tab1}.  }\end{figure}

In Fig.~\ref{mu1} the convergence of the $\mu^{\mathrm{ex}}$ computed with different methods is shown as a function of the MD time for the less dense $N=720$ system. Due to the relatively low density Widom method converges efficiently. Therefore we report the Widom estimate obtained with a large number of insertions ($M_{\mathrm{w}}=27000$), to be used as a reference value for the other calculations. We note that the Widom result at $M_{\mathrm{w}}=64$, and the metadynamics results at $M=64$ (type I and II), converge to the correct chemical potential, despite the larger statistical error due to the smaller number of insertions (see Tab.~\ref{tab1}). 

In Fig.~\ref{mu1} we also report the results of two TI calculations, TI-a and TI-b. The MD time indicates the overall time of sampling for all $\lambda_{i}$ values. This overall time does not include the thermalization time $T_{\mathrm{th}}=500\Delta\tau$ applied whenever $\lambda$ is adjusted to a new value (which represents a further computational cost). TI-a and TI-b are performed at the same conditions and computational setup, except for the initial configuration of the system. As shown in Tab.~\ref{tab1}, the resulting $\mu^{\mathrm{ex}}$ are non-consistent with the reference Widom value and with each other, despite being obtained with equivalent setup.
This is because TI relies on the sampling performed by the $N+1$-th atom at each $\lambda_{i}$ step, in order to compute the $\langle\partial U/\partial \lambda\rangle_{\lambda=\lambda_{i}}$ averages. Because of the non-homogeneity a thorough sampling of the $z$ coordinate is required, and this slows down the convergence of the averages. This poor convergence motivates the non-compatible results obtained by TI-a and TI-b. On the contrary, because of Eq.~\ref{widomnon}, both Metadynamics and Widom techniques do not suffer from this problem. For this reason, for the $N=920$ case, we restrict to these two techniques.

\begin{table}[]
\centering
\setlength{\tabcolsep}{10pt} 
\renewcommand{\arraystretch}{1.2} 
\begin{tabularx}{9cm}{{|C|C|}}
\hline
Method & $\mu^{\mathrm{ex}}\ [\epsilon]$\\ 
\hline
Widom $M_{\mathrm{W}}=27000$ & $-3.252\pm0.001$ \\ 
Widom $M_{\mathrm{W}}=64$ & $-3.241\pm0.017$ \\
Meta I $M=64$ & $-3.243\pm0.012$ \\
Meta II $M=64$ & $-3.241\pm0.012$ \\
TI-a & $-3.365\pm0.02$ \\
TI-b & $-3.315\pm0.019$ \\ 
\hline
\end{tabularx}
\caption{\label{tab1} $\mu^{\mathrm{ex}}$ estimates obtained for the $N=720$ system. The first column indicates the calculation method. The chemical potential is computed over $5\times10^{5}\Delta\tau$ MD steps. The uncertainty corresponds to the statistical error of the estimate, calculated combining bootstrap and block averaging.}
\end{table}

\begin{figure}[] 
\centering \includegraphics{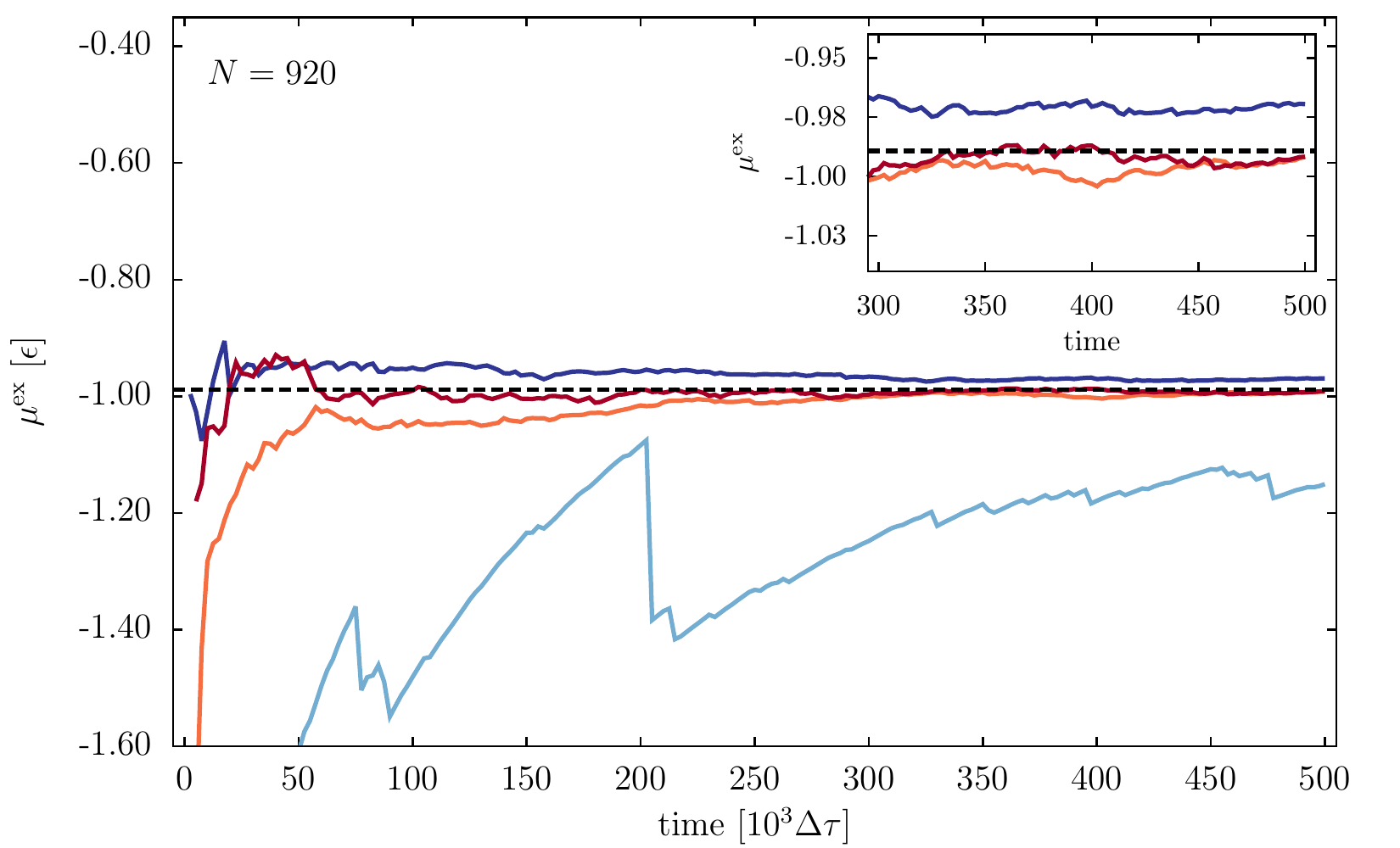} 
\caption{\label{mu2} $\mu^{\mathrm{ex}}$ as a function of the MD time for the $N=920$ system. The Widom calculations with $M_{\mathrm{w}}=27000$ (blue) and $M_{\mathrm{w}}=64$ (light blue) are compared to metadynamics results with $M=64$ of type I (red) and II (orange). The black dashed line indicates the $\mu^{\mathrm{ex}}$ value obtained with a $M=512$ metadynamics run (type I) of $5\times10^{5}\Delta\tau$. The inset shows the final $2\times10^{5}\Delta\tau$ steps, with increased $\mu^{\mathrm{ex}}$ resolution, in order to underline the systematic deviation between the estimates. The final results, and respective statistical errors, are indicated in Tab.~\ref{tab2}.}\end{figure}
\begin{table}[]
\centering
\setlength{\tabcolsep}{10pt} 
\renewcommand{\arraystretch}{1.2}
\begin{tabularx}{9cm}{{|C|C|}}
\hline
Method & $\mu^{\mathrm{ex}}\ [\epsilon]$\\ 
\hline
Widom $M_{\mathrm{W}}=27000$ & $-0.97\pm0.01$ \\ 
Widom $M_{\mathrm{W}}=64$ & $-1.15\pm0.19$ \\
Meta I $M=64$ & $-0.99\pm0.01$ \\
Meta II $M=64$ & $-0.99\pm0.01$ \\
Meta I $M=512$ & $-0.99\pm0.01$ \\
\hline
\end{tabularx}
\caption{\label{tab2} $\mu^{\mathrm{ex}}$ estimates obtained for the $N=920$ system. The first column indicates the calculation method. The chemical potential is computed after $5\times10^{5}\Delta\tau$ steps. The uncertainty corresponds to the statistical error of the estimate, calculated combining bootstrap and block averaging.}
\end{table}

In Fig.~\ref{mu2} we report the convergence of $\mu^{\mathrm{ex}}$ for the more dense $N=920$ system. We compare two Widom calculations, with $M_{\mathrm{w}}=27000$ and $M_{\mathrm{w}}=64$, and two metadynamics calculations, of type I and II, with $M=64$. 
The Widom calculation with $M_{\mathrm{w}}=64$ does not converge in the allotted time, while the other three methods converge in about the same timescale.
However, we note that the $M_{\mathrm{w}}=27000$ Widom estimate, even though based on a larger insertion statistics, has an error comparable to that of metadynamics. Most importantly, we note that both metadynamics estimates deviate from the apparently converged Widom result (see the inset of Fig.~\ref{mu2}). To assess the reason for this deviation we first estimated $\mu^{\mathrm{ex}}$ with a more expensive metadynamics simulation of type I, using $M=512$ insertion points. The larger statistics of insertions determines a more precise estimate. However, the result deviates in a similar way from the Widom value, and is compatible with the other two metadynamics estimates (see Tab.~\ref{tab2}).

As discussed in Ref.~\onlinecite{LuJCP2001a}, free-energy calculations such as this one are affected by a systematic error. 
\begin{figure}[] 
\centering\includegraphics{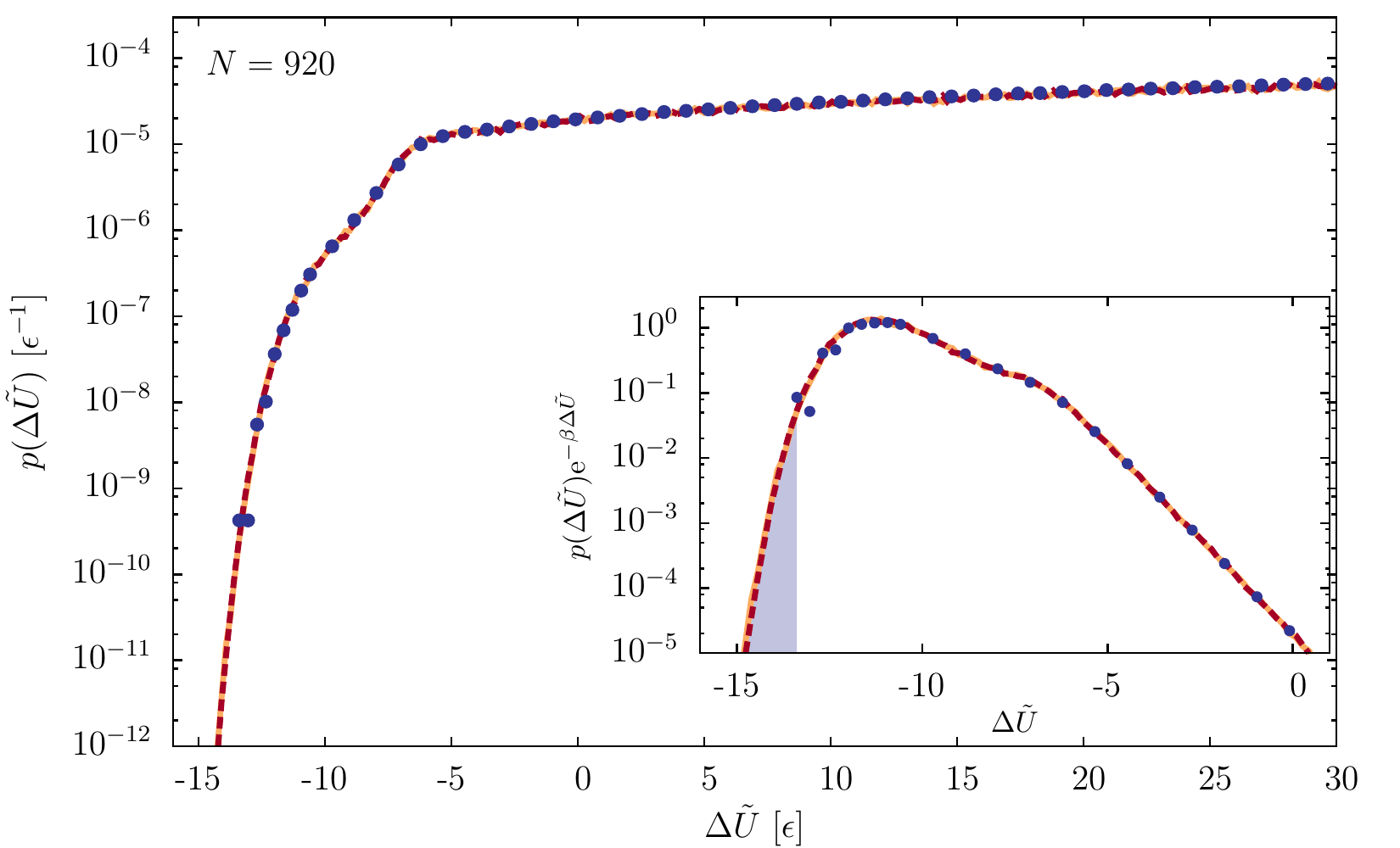}\\  
\caption{\label{pi} Behavior of $p(\Delta \tilde{U})$ for the $N=920$ system simulations. The Widom result with $M_{\mathrm{w}}=27000$ (blue dots) is compared to those of metadynamics runs with $M=64$, type I (orange solid curve) and type II (red dashed curve) calculations. The inset displays the integrand of Eq.~\ref{muofrho2}, namely $p(\Delta \tilde{U})\exp(-\beta\Delta\tilde{U})$, for the three calculations. The light blue shading highlights the energy region sampled only via metadynamics.}\end{figure}
This can be understood if we rewrite the excess chemical potential as a function of the effective insertion energy distribution, namely $p(\Delta \tilde{U})=\left\langle\delta[\Delta \tilde{U}-\Delta \tilde{U}(\mathbf{R}^{*};\mathbf{R})]\right\rangle$, obtaining:
\begin{equation}
\label{muofrho2}
\mu^{\mathrm{ex}}=-\beta^{-1}\log\int\mathrm{e}^{-\beta \Delta\tilde{U}}p(\Delta\tilde{U})\,\mathrm{d}\Delta\tilde{U}.
\end{equation}
Eq.~\ref{muofrho2} shows that the largest contributions to $\mu^{\mathrm{ex}}$ come from the negative tail of the insertion energy distribution. Thus, the insufficient sampling of this tail determines a systematic deviation in the estimate of the chemical potential. This is shown in Fig.~\ref{pi}, where we contrast the $p(\Delta\tilde{U})$ obtained in the Widom calculation with those resulting from the two metadynamics runs. In Widom method the negative values of $\Delta\tilde{U}$ are not sufficiently sampled. In the other two calculations, thanks to metadynamics ability to enhance rare fluctuations, the negative $\Delta\tilde{U}$ tail is thoroughly sampled and, as a result, the systematic error is rendered negligible.

We can thus conclude that in this density regime our metadynamics approach does not suffer from the inaccuracy affecting the unbiased insertion sampling, providing more reliable estimates than the Widom method. 

\section{Conclusions}\label{conclusions}

In this work we have presented a development of an enhanced sampling method for computing the excess chemical potential of dense fluids, recently published in Ref.\cite{PeregoEPJ2016}. The same approach has been here generalized to non-homogeneous fluids by the definition of a new, more general CV that accounts for the density variations over the system volume.
We have tested this method with a non-homogeneous LJ fluid, at two different density regimes, and compared our results with Widom and TI calculations. In both cases the convergence of our method has shown to be competitive with the reference techniques, and particularly advantageous in the high density regime.
When a dense fluid is considered, the enhanced sampling simulations require a smaller statistics of insertion to reach the precision of Widom method. On top of that, by thoroughly sampling the negative tail of the insertion energy distribution, our technique also prevents the occurrence of a systematic error in the final estimate, providing a more accurate chemical potential value. 

The calculations presented here were performed by using WT metadynamics to enhance the sampling. However, other biasing techniques can be usefully employed, as e.g.~the recently proposed VES method. As an example we have compared the use of these two sampling techniques, showing that the flexibility of the latter can boost the performances of the calculation (the results are reported in the SM). 
The extension of the technique to non-homogeneous liquids is a crucial step for the method to be employed on more realistic systems, especially in the study of phase coexistence and nano-confined liquids. 

\section*{Supplementary Material}
See supplementary material for further details on the accuracy of the presented method and for the results of chemical potential calculations using VES as enhanced sampling technique.

\begin{acknowledgements}
The authors thank M.~Heidari, R.~Potestio and J.~Smrek for a critical reading of the manuscript. The computational resources were provided by the Swiss Center for Scientific Computing. The authors acknowledge research funding through the National Center for Computational Design and Discovery of Novel Materials MARVEL and the VARMET European Union Grant ERC-2014-ADG-670227.
\end{acknowledgements}

\bibliography{bibfile}

\end{document}